\documentstyle[11pt,a4]{article}

\begin{document}

\title {\bf Hyperbolic theories of dissipation: Why and when do we need them?}

\author{Luis Herrera\thanks{Postal address: Apartado 80793, Caracas 1080A,
Venezuela; E-mail address: laherrera@telcel.net.ve}\\ Escuela de
F\'{\i}sica, Facultad de Ciencias,\\ Universidad Central de Venezuela,
Caracas, Venezuela.\\ and\\ Diego Pav\'{o}n\thanks{E-mail address:
diego@ulises.uab.es}\\ Departamento de F\'{\i}sica, Facultad de Ciencias,\\
Edificio Cc, Universidad Aut\'{o}noma de Barcelona,\\ 08193 Bellaterra, Spain.
}

\date{}
\maketitle

\begin{abstract}
We illustrate and emphasize the relevance of hyperbolic theories of dissipation
in different physical scenarios. Particular attention is paid to
self--gravitating systems where the relaxation time may become large enough
as to require a description of the transient regime. It is argued that even
outside that regime, hyperbolic theories may be needed to
provide an accurate description of dissipative processes.
\end{abstract}

\newpage

\section{Introduction}

In the description of the evolution of any physical system, it is mandatory
to evaluate, as accurateely as possible, the order of magnitude of different
characteristic time scales, since their relationship with the time scale
of observation (the time during which we assume our  description of the
system to be valid) will determine, along with the relevant  equations, the
evolution pattern. Take a forced damped harmonic oscillator and consider its
motion on a time scale much larger than both the damping time and the period
of the forced oscillation. Then, what one observes is just a harmonic motion.
Had we observed the system on a time scale of the order of (or smaller) than
the damping time, the transient regime would have become apparent. This
is rather general and of a very relevant interest when dealing with
dissipative systems. It is our purpose here, by means of examples and
arguments related to a wide class of phenomena, to emphasize the convenience
of resorting to hyperbolic theories when dissipative processes, either
outside the steady--state regime or when the observation time is of the
order or shorther than some characteristic time of the system, are under
consideration. Furthermore, as it will be mentioned below, transient
phenomena may affect the way in which the system leaves the equilibrium,
thereby affecting  the future of the system even for time scales much
larger than the relaxation time.

\section{The rationale behind hyperbolic theories}

Parabolic theories of dissipative phenomena have long and a venerable history
and proved very useful especially in the steady--state regime \cite{Fetter}.
They exhibit however some undesirable features, such as acausality (see
e.g., \cite{Hiscock}, \cite{Jou}), that prompted the formulation  of hyperbolic
theories of dissipation to get rid of them \cite{Muller}, \cite{Jou}. This was
achieved at the price of extending the set of field variables by including
the dissipative fluxes (heat current, non--equilibrium stresses and so on) at
the same footing as the classical ones (energy densities, equilibrium
pressures, etc), thereby giving rise to a set of more physically satisfactory
(as they much better conform with experiments) but involved theories from the
mathematical point of view. These theories have the additional advantage of
being backed by statistical fluctuation theory, kinetic theory of gases
(Grad's 13--moment approximation), information theory and correlated
random walks (at least in the version of Jou {\it et al.}) \cite{Jou}.

A key quantity in these theories is the relaxation time $\tau$ of the
corresponding  dissipative process. This positive--definite quantity has a
distinct physical meaning, namely the time taken by the system to return
spontaneously to the steady state (whether of thermodynamic equilibrium or
not) after it has been suddenly removed from it. It is, however,  somehow
connected to the mean collision time $t_{c}$ of the  particles responsible
for the dissipative process, ofentimes erroneously identified with it.
In principle they are different since $\tau$ is (conceptually
and many times in practice) a macroscopic time, although in some instances
it may correspond just to a few $t_{c}$. No general formula linking
$\tau$ and $t_{c}$ exists, their relationship depends in
each case on the system under consideration. As mentioned above,
it is therefore appropriate to interpret $\tau$ as the time taken by
the corresponding dissipative flow to relax to its steady value.

Thus, it is well known that the classical Fourier law for heat current,
\\
\begin{equation}
\vec{q}  = - \kappa \vec{\nabla}T \, ,
\label{q}
\end{equation}
\\
with $\kappa$ the heat conductivity of the fluid, leads to a parabolic
equation for temperature (diffusion equation),
\\
\begin{equation}
\quad\frac{\partial T}{\partial t} = \chi \nabla^2T  \, \qquad \qquad
(\chi \equiv \frac{\kappa}{\rho \, c_{p}})
\label{dif}
\end{equation}
\\
(where $\chi$, $\rho$ and $c_{p}$ are the diffusivity, density and specific
heat at constant pressure, respectively), which
does not forecast propagation of perturbations along characteristic causal
light--cones (see \cite{Jou}, \cite{Joseph}, \cite{Maartens}, \cite{Tzou} and
references therein). That is to say, perturbations propagate with infinite
speed. This non--causal behavior is easily visualized by taking a look at
the thermal conduction in an infinite one dimensional medium (see e.g.
\cite{Fetter}, \cite{Landau}). Assuming that the temperature of the line
is zero for $t<0$, and putting a heat source at $x=x_{0}$ when $t=0$, the
temperature profile for $t > 0$ is given by
\\
\begin{equation}
T \propto \frac{1}{\sqrt t} \exp{\left[-
\frac{(x-x_0)^2}{t}\right]} \, ,
\label{infinite}
\end{equation}
\\
implying that
for $t=0 \; \Longrightarrow T= \delta (x-x_{0})$, and for
$t=\tilde t>0 \; \Longrightarrow  T \neq 0 \; \forall x $.

\noindent
In other words, the presence of a heat source at $x_{0}$ is
instantaneously felt by all observers on the line, no matter how
far away from $x_{0}$ they happen to be. The origin of this behavior
can be traced to the parabolic character of Fourier's law, which
implies that the heat flow starts (vanishes) simultaneously with
the appearance (disappearance) of a temperature gradient. Although
$\tau$ is very small for phonon--electron, and phonon--phonon interaction
at room temperature (${\cal O}(10^{-11})$ and ${\cal O}(10^{-13})$ seconds,
respectively \cite{Peierls}), neglecting it is the source of difficulties,
and in some cases a bad approximation as for example in superfluid Helium
\cite{Peshkov}, and degenerate stars where thermal conduction is dominated
by electrons -see \cite{Jou}, \cite{Joseph}, \cite{mdv}, for further
examples.

In order to overcome this problem Cattaneo and (independently) Vernotte
by using the relaxation time approximation to Boltzmann equation for a
simple gas derived a generalization of Fourier's law, namely
\cite{Cattaneo}
\\
\begin{equation}
\tau \frac{\partial \vec q}{\partial t} + \vec q = - \kappa \vec \nabla T.
\label{Cattaneo}
\end{equation}
\\
This expression (known as Cattaneo-Vernotte's equation) leads to a
hyperbolic equation for the temperature (telegraph equation)
\\
\begin{equation}
\tau \frac{\partial^2 T}{\partial t^2 } + \frac{\partial T}{\partial t}
=  \chi \nabla^{2} T \, ,
\label{telegraph}
\end{equation}
\\
which describes the propagation of thermal signals with a finite speed
\\
\begin{equation}
v = \sqrt{\chi/ \tau}.
\label{velocity}
\end{equation}
\\
This diverges only if the unphysical asumption of setting $\tau$ to
zero is made.

It is worth mentioning that a simple random walk analysis of transport
processes naturally leads to telegraph equation, not to the diffusion
equation  -see e.g. \cite{weymann}. Again, the latter is obtained only
if one neglects the second derivative term.

It is instructive to write (\ref{Cattaneo}) in the equivalent
integral form
\\
\begin{equation}
\vec q = - \frac{\kappa}{\tau} \int_{-\infty}^{t}{\exp{\left[-
\frac{(t-t')}{\tau}\right]} \cdot \vec \nabla T(\vec x,t') dt'},
\label{catanneointegral}
\end{equation}
\\
which in turn is a particular case of the more general expression
\\
\begin{equation}
\vec q = - \int^{t}_{-\infty}{Q(t-t') \vec \nabla T(\vec x,t') dt'}.
\label{generalintegral}
\end{equation}
\\
The physical meaning of the kernel $Q(t-t')$ becomes obvious by
observing that
\\
\begin{equation}
\begin{array}{cccc}
{\rm for}  & Q = \kappa \delta(t-t')  &  \Longrightarrow   &  \vec q =
- \kappa \vec \nabla T \quad {\rm (Fourier)}
\\
{\rm for}  &  Q =  {\rm constant}     &   \Longrightarrow   &   \frac{
\partial^2 T}{\partial t^2} = \chi \nabla^{2} T,
\end{array}
\end{equation}
\\
i.e., $Q$ describes the thermal memory of the material by assigning
different weights to temperature gradients at different moments in the past.
The Fourier law corresponds to a zero--memory material (the only relevant
temperature gradient is the ``last" one, i.e., the one simultaneous with
the appearance of $\vec{q}$). By contrast the  infinite memory case (with 
$Q=$ constant) leads to an undamped wave. Somewhere in the middle is the
Cattaneo-Vernotte equation, for which all temperature gradients
contribute to $\vec{q}$, but their relevance goes down as we
move to the past.

From these comments it should be clear that different classes of dissipative
systems may be described by different kernels. The one corresponding to
(\ref{Cattaneo}) being suitable for the description of a restricted
subclass of phenomena.

Obviously, when studying transient regimes, i.e., the evolution from a 
steady--state situation to a new one,  $\tau$ cannot be neglected. In 
fact, leaving aside that parabolic theories are necessarily non--causal,
it is obvious that whenever the time scale of the problem under
consideration becomes of the order of (or smaller) than the relaxation time,
the latter cannot be ignored. It is common sense what is at stake here:
neglecting the relaxation time ammounts -in this situation- to
disregarding the whole problem under consideration.

\section{Hyperbolic versus parabolic}
According to a basic assumption underlying the disposal of hyperbolic
dissipative theories, dissipative processes with relaxation times
comparable  to the characteristic time of the system are out of the
hydrodynamic regime.  However, the concept of hydrodynamic regime
involves the ratio between the  mean free path of fluid particles  and
the characteristic length of the system. When this ratio is lower that
unity, the fluid is within the hydrodynamic regime. When it is larger
than unity, the regime becomes Knudsen's. In the latter case the fluid is
no longer a continuum and even hyperbolic theories cease to be realiable.

Therefore that assumption  can be valid only if the particles making up the
fluid are the same ones that transport the heat. However, this is (almost?)
never the case. Specifically, for a neutron star, $\tau$ is of the
order of the scattering time between electrons (which carry the
heat) but this fact is not an obstacle (no matter how large the
mean free path of these electrons may be) to consider the neutron
star as formed by a Fermi fluid of degenerate neutrons. The same
is true for the second sound in superfluid Helium and solids, and
for almost any ordinary fluid. In brief, the hydrodynamic regime
refers to fluid particles that not necessarily (and as a matter of fact,
almost never) transport the heat. Therefore large relaxation times (large
mean free paths of particles involved in heat transport) does not imply a
departure from the hydrodynamic regime (this fact has been streseed before
\cite{Santos}, but it is usually overlooked).

However, even in the case that the particles making up the fluid are
responsible of the dissipative process, it is not ``always" valid to take for
granted that $\tau$ and $t_{c}$ are of the same order  (se e.g. \cite{Geroch},
\cite{Lindblom}), or what comes to the same that the dimensionless quantity
$\Gamma \equiv (\tau c_{s}/L)^{2}$ is negligible in all instances -here
$c_{s}$ stands for the adiabatic speed of sound in the fluid under
consideration and $L$ the characteristic length of  the system. That
assumption would be right if $\tau$ were always comparable to $t_{c}$ and
$L$ always ``large", but there are, however, important situations in which
$\tau \gg t_{c}$, and  $L$ ``small" although still large enough to
justify a macroscopic description. For tiny semiconductor pieces of
about $10^{-4}$ cm in size, used in common electronic
devices submitted to high electric fields, the
above dimensionless combination (with $\tau \sim 10^{-10}$
sec, $c_{s} \sim 10^{7}$ cm/sec \cite{muscato}) can easily
be of the order of unity. In ultrasound propagation as well as
light-scattering experiments in gases and  neutron-scattering in
liquids the relevant length is no longer the system size,
but the wavelenght $\lambda$ which is usually much
smaller than $L$ \cite {Weiss}, \cite{Copley}.
Because of this, hyperbolic theories may bear some
importance in the study of nanoparticles and quantum dots.
Likewise in polymeric fluids relaxation
times are related to the internal configurational
degres of freedom and so much longer than $t_{c}$
(in fact they are in the range of the minutes), and
$c_{s} \sim 10^{5}$ cm/sec, thereby $\Gamma \sim {\cal O}(1)$.
In the degenerate core of aged stars the
thermal relaxation time can be as high as $1$ second
\cite{Harwit}. Assuming the radius of the core of about
$10^{-2}$ times the solar radius, one has $\Gamma \sim {\cal O}(1)$
again. Fully ionized plasmas exhibit a collisionless regime
(Vlasov regime) for which the parabolic hydrodynamics predicts
a plasmon dispersion relation at variance with the microscopic results;
the latter agree, however, with the hyperbolic hydrodynamic approach
\cite{Tokatly}. Think for instance of some syrup fluid flowing
under a imposed shear stress, and imagine that the shear is suddenly
switched off. This liquid will come to a halt only after a much longer
time ($\tau$) than the collision time between its constituent
particles has elapsed.

The fact that $\tau$ can quantitatively greatly differ from $t_{c}$ is most
dramatically suggested by the matter--radiation decoupling in the
early universe. In a recent paper Pav\'{o}n and Sussman \cite{roberto}
by using the Lemaitre--Tolmann--Bondi metric \cite{LTB} along with the
hyperbolic transport equation for the shear--stress
\\
\begin{equation}
\tau \dot\Pi_{cd}\,h^c_ah^d_b+\Pi_{ab}\left[
1+{\textstyle{1\over{2}}}T\eta\left({{\tau}\over{T\eta}}  \,
u^c\right)_{;c}\right]+ 2\eta\,\sigma_{ab}= 0, \quad \mbox {with}
\quad
\eta = {\textstyle{4\over{5}}}p \tau \, ,
\label{pitensor}
\end{equation}
\\
showed that the relaxation time of shear viscosity results several ordersof magnitude larger than the Thomson collision time between photons and
electrons for most of the radiative era, i.e., in the temperature range
$ 10^{3} < T < 10^{6}$ Kelvin. (In above expression $\Pi_{ab}$ denotes
the shear--stress tensor arising from the matter--radiation interaction,
$\eta$ is the transport coefficient, and $\sigma_{ab}$ the symmetric
trace--free part of the gradient of the four--velocity, and $p$
the radiation pressure. In this scenario $t_{c}$ is obtained by
introducing the Thomson cross--section in Saha's formula.

Many other examples could be added but we do not mean to be exhaustive.

Even in the steady regime the descriptions offered by parabolic and
hyperbolic theories do not necessarily coincide. The differences
between them in such a situation arise from (i) the presence of
$\tau$ in terms that couple the vorticity to the heat flux and
shear stresses. These may be large even in steady states (e.g.
rotating stars). There are also other acceleration coupling terms
to bulk and shear stresses and heat flux. The coefficientsfor these vanish
in parabolic theories, and they could
be large even in the steady state. (ii) From the convective
part of the time derivative (which are not negligible
in the presence of large spatial gradients). (iii) From
modifications in the equations of state due to the presence of
dissipative fluxes \cite{Jou}.

However, it is precisely before the establishment of the steady
regime that both types of theories differ more importantly. It
is well--known (see  \cite{Jou}, \cite{Joseph}, \cite{mdv},
\cite{Herrera}) that a variety of physical processes
take place on time scales of the order of (or even smaller) than
the corresponding relaxation time, which as was stressed above
does not imply that the system is out of hydrodynamic regime.
Therefore if one wishes to study a dissipative process for times
shorter than $\tau$, it is mandatory to resort to a hyperbolic
theory which is a more accurate macroscopic approximation
to the underlying kinetic description.

In the particular case of neutron star matter, it appears that the
relaxation time may indeed be of the order of magnitude of the
characteristic time of some physical relevant processes. Thus,
from  (\ref{velocity}) one has
\\
\begin{equation}
\tau = \frac{\kappa}{v^{2} \, \rho \, c_{p}}.
\label{tau}
\end{equation}
\\
If the heat conductivity is dominated by electrons (as is the case of
a neutron star), then we can adopt the expression \cite{Flowers}
\\
\begin{equation}
\kappa \approx 10^{23}[\rho/10^{14} \mbox{g \, cm}^{-3}]
[10^{8} K/T] \mbox{erg \, s}^{-1} \, \mbox{cm}^{-1} \, \mbox{K}^{-1}
\label{kap}
\end{equation}
\\
and for the specific heat $c_{p} = \beta T/M$, where $\beta$
(which is model dependent) is given by \cite{Shibazaki}
$\beta \approx 10^{29}$ erg/K$^{2}$ and $M$ denotes the total mass.
Feeding back these expressions  in (\ref{tau}) we get
\\
\begin{equation}
\tau \approx \frac{10^{20}}{\left[T^2\right] \left[v^2\right]} \; \mbox{s},
\label{taub}
\end{equation}
\\
where $[T]$ and $[v]$ denote the numerical values of the temperature and
the velocity of the thermal wave in Kelvin and cm/s, respectively.
We have further assumed $\rho \approx 10^{14} \mbox{g \, cm}^{-3}$ and
approximately $10 \, $ Km  for the radius of the degenerate core.
In this case, assuming for $v$ the speed of light in vacuum, for
temperatures of the order of $[T] \approx 10^{2}$ Kelvin -a certainly
low value that corresponds to the latest phases in the evolution
of a neutron star  (see \cite{Shibazaki})- we get
$\tau \approx 10^{-5}$ seconds.

\noindent
However, for more reasonable values of $v$, such as $ 10^{3}$ cm/s,
corresponding to the speed of second sound in superfluid helium,
and for temperatures of the order of $10^9$ Kelvin, one obtains
$\tau \approx 10^{-4}$ seconds. Whereas for $T \approx 10^{6}$ Kelvin, the
relaxation time may be found as large as $\tau \approx 10^{2}$ seconds.
Therefore physical process on a time scale of the order (or smaller) than
above belong to the transient regime and require a hyperbolic theory.

Another astrophysical phenomenom where the effects of the transient
regime can be felt is the double peaked temporal luminosity profile
observed from x--ray bursters in star explosions, such as the one
reported by Hoffman {\it et al.} \cite{Hoffman}. These authors observed
a x--ray burster showing a precursor peak, that lasted about 4 seconds, neatly
separated from the main event -which lasted above 1000 seconds. While
the analysis of elastic photon diffusion through a cloud of ionized
plasma of modest optical depth $\tau_{*}$ via parabolic theories just
predicts the convential diffusion structure, the hyperbolic theory predicts
a well defined peak for times shorter than $ t_{D} \sim \tau_{*}^{2}t_{c}/2$
and a diffusive behavior for $t > t_{D}$ \cite{Schweizer}. This kind of
event is reminiscent of laser induced heat pulses in solids where the double
structure is evident \cite{Dreyer}. Again hyperbolic theories successfully
deal with such situations while parabolic do not \cite{Jou}.

Finally, it is worth mentioning that relaxation time appears to be quite
relevant in the outcome of gravitational collapse as many numerical and
analytical calculations indicate (see  \cite{Herrera}, \cite{Govender} and
references therein). In particular it has been shown that,  for otherwise
identical boundary and initial conditions, larger values of $\tau$
imply more flattened  and long lasting pulses of emission, thereby
affecting the evolution of the object.

\section{Beyond the transient regime}
So far we have seen that only for times larger than $\tau$ it is sensible
to resort to a parabolic theory (modulo that the spatial gradients are not
so large that the convective part of the time derivative becomes important,
and that the fluxes and coupling terms remain safely small).
However, even in these cases, it should be kept in mind that the way a
system leaves the equilibrium may be very sensitive to the relaxation
time.

Indeed, it has been shown \cite{eim} that  after the fluid leaves the
equilibrium, on a time scale of the order of relaxation time, the 
effective inertial mass density of a dissipative fluid reduces by a 
factor that depends on the dissipative variables. By ``effective inertial 
mass density" (EIMD) we mean the factor of proportionality between the 
applied three--force density and the corresponding proper acceleration 
(i.e., the three--acceleration measured in the instantaneous rest frame). 
The expression for the EIMD contains a contribution from dissipative 
variables which reduces its value with respect to the 
equilibrium situation.

Specifically, it has been shown that just after leaving the equilibrium 
on a time scale of the order of $\tau$ the EIMD  becomes (in 
relativistic units)
\begin{equation}
\Lambda = (\rho+p) ({1-\alpha}),
\label{nueva}
\end{equation}
\\
with
\begin{equation}
\alpha = \frac{\kappa T}{\tau (\rho + p)} \, ,
\label{alpha}
\end{equation}
\\
giving rise, in principle, to the possibility of a vanishing $\Lambda$
(i.e., $\alpha = 1$) or even negative ($\alpha > 1$). Such effect may be
present in any dissipative fluid (either self--gravitating or not)
\cite{tetra}.

In order to evaluate $\alpha$, let us turn back to coventional units.
Assuming for simplicity $\rho + p \approx 2\rho$, we obtain
\\
\begin{equation}
\alpha = \frac{\kappa T}{\tau (\rho + p)}
\approx \frac{[\kappa][T]}{[\tau][\rho]} \times 10^{-42},
\label{un}
\end{equation}
\\
where $[\kappa]$, $[T]$, $[\tau]$, $[\rho]$ denote the numerical
values of these quantities in c.g.s units.

Obviously, except for extremely high values of $\kappa$ and $T$,
$\alpha$ will be a much less than unity. Observe, in this connection,
that although the smaller $\tau$, the larger $\alpha$, this is of little
consequence since (\ref{nueva}) is valid only on a time scale of the order of
$\tau$. Therefore, a decreasing of $\Lambda$ with physically relevant impact
requires values of (\ref{un}) close to unity, due to large values of $\kappa$
and $T$ but non--negligible values of $\tau$.

A possible scenario where $\alpha$ may decrease substantially (for
non-negligible values of $\tau$) might be provided by a pre-supernova
event. At the last stages of massive star evolution, the decreasing of
the opacity of the fluid, from very high values preventing the
diffusion of photons and neutrinos (trapping \cite{Ar}), to smaller
values, gives rise to a radiative heat conduction. Upon these conditions
both $\kappa$ and $T$ could become sufficiently large as to imply a
substantial increase of $\alpha$. Indeed, the values suggested in
\cite{Ma} ($[\kappa] \approx 10^{37}$;
$[T] \approx 10^{13}$; $[\tau] \approx 10^{-4}$; $[\rho] \approx
10^{12}$, in c.g.s. units) lead to $\alpha \sim {\cal O}(1)$. The
obvious consequence would be to enhance the efficiency of whatever
expansion mechanism, of the central core, because of the decreasing
of $\Lambda$. Thus, it becomes clear that the value of $\alpha$ in the
transient regime may critically affect the subsequent evolution of the
system. Therefore it may be said that the future of the system at time
scales much longer than the relaxation time (once the steady state is
reached), may also critically depend on $\tau$.

\section{Concluding remarks}
It has been argued that the infinite speeds of propagation predicted by
parabolic theories is just a consequence of using them outside their
range of applicability which is limited by the discrete nature of matter
\cite{weymann}. In this connection it should be noted that hyperbolic
theories also share this constraint but they naturally predict speeds
in good agreement with experiments. Therefore one is led to conclude
that hyperbolic theories have a range of aplicability much wider than
parabolic theories.

Aside from $\tau$ hyperbolic theories introduce a certain number of new
parameters that couple the different dissipative fluxes. In spite of this
being rather natural (as these theories are designed to explain more
complex phenomena than parabolic theories do), this feature has been
under unduly criticism as though these quantities were ``free parameters"
that one could choose at will \cite{Geroch}. This is not the case. On the
one hand, their number gets severely reduced when one realizes that
they are interrelated. For instance, for an ideal gas under heat flux
and bulk and shear stresses the parameters entering the transport
equations of second order hyperbolic theory (equations (2.31)--(2.33)
in [3.a]) are linked by the six equations (3.44)--(3.45) in [3.a].
On the other hand, these parameters are restricted by the convexity of
the entropy function, and can be calculated explicitly with the help of
kinetic theory, or fluctuation theory, such as in the case of a
radiative fluid \cite{radiative}.

In summary, although parabolic theories have proved very useful for many
practical purposes, they  appear to fail hopelessly in a number of well--known
instances (such as transient regimes). By contrast, hyperbolic theories
successfully predict the experimental results, and so, in these regimes
hyperbolic theories happen to be more reliable. In the steady--state
(under the conditions mentioned above) and for times exceeding $\tau$
both theories converge. Before closing we would like to emphasize we are
not advocating to stop using parabolic transport equations. Our aim is
simply to stress the convenience of using hyperbolic transport equations
there and when parabolic theories either fail or the problem under
consideration happens to lie outside their range of applicability.

We hope this paper will help to convince the reader that hyperbolic
theories are indeed of not mere academic interest and it would not be
wise to dispense of them.

\section*{Acknowledgements}
The authors are indebted to David Jou for reading the manuscript and
helpful remarks. This work has been partially supported by the Spanish
Ministry of Science and Technology under grants BFM 2000-C-03-01 and
2000-1322.

\end{document}